\documentclass{jaa}
\usepackage{graphicx}
\usepackage{gensymb}
\usepackage{authblk}
\begin{document}\sloppy

%%paper title
%%For line breaks \\ can be used within title 
\title{Observations of bright stars with $AstroSat$ Soft X-ray Telescope.} 
%\\ Second line}

%%author names are separated by comma (,) 
%%use \and before the last author name 
%%use a * along with the number separated by comma
%% for the  author for correspondence
%%\textsuperscript{number} is used for affiliation
%%\affilOne, \affilTwo etc., upto \affilTwentyfive is possible
%%Please note the first letter after \affil is capitalised in the command
%%

\author{K. P. Singh\textsuperscript{1,*},  G. Stewart \textsuperscript{2}, S. Chandra \textsuperscript{3}, G. C. Dewangan \textsuperscript{4}, S. Bhattacharyya\textsuperscript{5}, N. S. Kamble\textsuperscript{5}, S. Vishwakarma\textsuperscript{5}, and J. G. Koyande\textsuperscript{5}}
\affilOne{\textsuperscript{1}Indian Institute of Science Education and Research Mohali, SAS Nagar, Sector 81,
P.O.Manauli- 140306, India.\\}
\affilTwo{\textsuperscript{2} Department of Physics and Astronomy, The University of Leicester, University Road, Leicester, LE1 7RH, UK\\}
\affilThree{\textsuperscript{3} Center for Space Research, North-West University, Potchefstroom, 2520, South Africa\\}
\affilFour{\textsuperscript{4} Inter-University Centre for Astronomy and Astrophysics, Ganeshkhind, Pune\\}
\affilFive{\textsuperscript{5} Department of Astronomy and Astrophysics, Tata Institute of Fundamental Research, Homi Bhabha Road, Mumbai 400005}

%%escape two column mode for title, affiliation and abstract
%%by giving \twocolumn command as shown

\twocolumn[{

\maketitle

%%include \corres to print the correspponding author Email id
\corres{kpsinghx52@gmail.com}

%%include \msinfo for
%%manuscript information such as
%%received, revised and accepted dates
%%
\msinfo{xxxx}{xxxx}

%%abstract
\begin{abstract}
We present observations of four bright stars observed with the $AstroSat$ Soft X-ray Telescope (SXT). 
Visible light from bright stars like these can leak through the very thin filter in front of the CCD in 
the focal plane CCD camera of the SXT and thus making the extraction of X-ray events difficult. Here, we show how
to extract the X-ray events without contamination by the visible light.  
The procedure applied to four bright stars here demonstrates how reliable X-ray information can be derived 
in such cases. The sample of bright stars studied here consists of two A spectral types (HIP 19265, HIP 88580), 
one G/K Giant (Capella), and a nearby M-type dwarf (HIP 23309). No X-ray emission
is observed from the A-type stars, as expected. X-ray spectra of Capella and HIP 23309 are derived and modeled here, 
and compared with the previous X-ray observations of these stars to show the reliability of the method used. 
We find that optical light can start to leak in the very soft energy bands below 0.5 keV for stars with V=8 mag.
In the process, we present the first X-ray spectrum of HIP 23309. 

\end{abstract}
%%insert keywords separated by 3 hyphens using \keywords{words}

}]
%%close the twocolumn escape here

%%include \doinum{number}for the DOI number in the header
%%include \volnum{number} for the volume number in the header
%%include \year{yyyy} for  year of publication in the header
%%include \pgrange{num--num} page range of article in the header
%%include \artcitid{num} for the article citation id
%%include \lp to print last page of the article
%%include \setcounter{page}{pagenum} for the exact starting page of the article

\doinum{12.3456/s78910-011-012-3}
\artcitid{\#\#\#\#}
\volnum{000}
\year{0000}
\pgrange{1--}
\setcounter{page}{1}
\lp{1}

\section{Introduction}
The focal plane camera in the Soft X-ray Telescope (SXT) (Singh et al. 2016, 2017) aboard the $AstroSat$ (Singh et al. 2014) carries 
a very thin optical light blocking filter in front of the CCD to block visible light but to allow the transmission of soft X-rays.   
The filter consists of a single fixed polyimide film which is 1840 Angstroms thick and coated with 488 Angstroms of 
Aluminum on one side. The filter is similar to the thin filter aboard the European Photon Imaging Camera 
(EPIC) (Struder et al. 2001) used in the {\it XMM-Newton} and the X-ray Telescope (XRT) aboard the $Swift$ Observatory (Burrows et al. 2005). 
The CCD used in the SXT is identical to the one used in the cameras of {\it XMM-Newton} and $Swift$.
The filter has to be thin to allow the transmission of soft X-rays while blocking the visible light 
from the cosmic X-ray sources. The X-ray transmission of the filter is shown in Figure 1.
The typical optical transmission of the filter is less than 5$\times$10$^{-3}$ 
(similar to the {\it XMM-Newton} thin filter and the filter onboard $Swift$ X-ray Telescope).
The filter design provides $\sim$7 magnitude of optical extinction over the visible band. 
For the $Swift$ XRT with a PSF of $\sim$15 arcsec a 6th magnitude star gives an optical loading of a few e- per pixel, 
at which point the quality of the X-ray data begins to be affected. 
For the SXT with a $\sim$7-8 times larger PSF and $\sim$2 times larger angular size of the pixel the 
safe optical limit is expected to be closer to a $\sim$4th magnitude star, but is needed to be verified by
post-launch observations and to check when the visible light can start leaking through the filter. 
This is specially important, as the SXT is occasionally pointed towards very bright stars with V$\leq$8 which 
can have a significant contribution to the events registered in the CCD due to visible light photons, 
thus contaminating the X-ray data in the very soft bands. For this purpose, we describe here SXT observations of a 
few bright stars two of which are non-X-ray emitting stars and the other two are bright X-ray emitting stars. 
We describe how to handle data from such observations and to obtain reliable X-ray information.
 
\begin{figure}[!thbp]
%Single column figure
\includegraphics[width=\columnwidth]{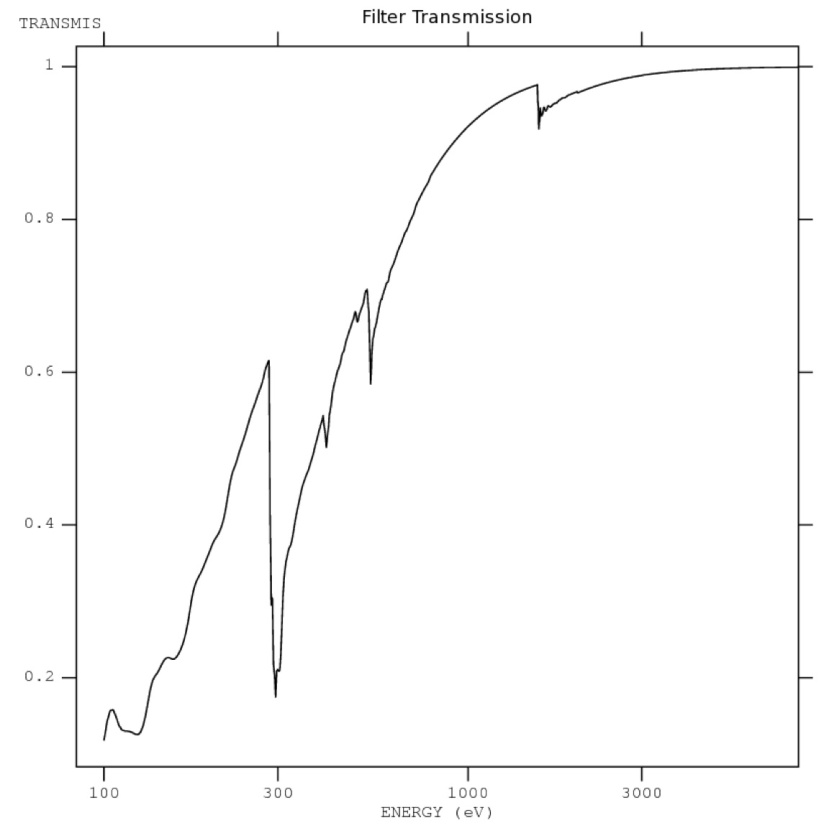}
\caption{X-ray transmission efficiency through the thin optical blocking filter in the SXT}
\label{figOne}
\end{figure}

\vspace{-1em}
\section{The Sample of Stars}

\begin{table*}[htb]
%\begin{table}
\small
%\tabularfont
%\centering
\caption{Properties of stars observed with \emph{SXT}}
\label{tab:tab1}
\begin{tabular}{llllll} 
\hline\hline
 Names & HIP 19265 & HIP 88580 & Capella  & HIP 23309 \\ 
 Other Names & HD 24716 & HD 165505 & $\alpha$Aur & CD-57 1054 \\ [0.5ex] 
\hline
Parameters\\
\hline
\textbf{Spectral Type} & A0 & A0   & G3 III & M0Ve \\
\textbf{B (mag)}   & 8.07   & 8.05 & 0.88   &  11.36 \\
\textbf{V (mag)}   & 7.94   & 7.96 & 0.08   &  9.98 \\
\textbf{R/G* (mag)}   & 7.94*  & 7.96*  & -0.52  &  9.89 \\
\textbf{Distance(pc)} & 325 & 226.6 & 13.1  & 26.90\\ 
\hline
\end{tabular}
\tablenotes{Distances are based on parallax measurements given in {\it Gaia} DR2 Bailer-Jones et al (2018), and at {\it Gaia} website:{gaia.ari.uni-heidelberg.de/tap.html}.}
\end{table*}

We have selected four bright single stars: HIP 19265, HIP 88580, Capella and HIP 23309 for our study here.  
Some of their important properties are listed in Table 1. 
Two of these have A0 spectral type which are generally known to be X-ray dark, and have 
never been detected in X-rays.  Stars with A-type spectrum are X-ray dark because they 
neither have an active corona or strong colliding and shocked winds to produce X-ray 
emission - the two processes known to produce X-rays in stars.  
A very small number of A type stars that have been detected in X-rays are all suspected to 
harbor a late type companion and thus not single or they have a very pecuilar chemistry 
or/and magnetic field (Ap or Am stars). 
One of the stars in our sample is Capella, a nearby G type giant that is known to be highly 
coronally active with copious X-ray emission and has been studied extensively in the past. 
Finally, we have an active M-type dwarf which was detected in the ROSAT All Sky Survey and 
has not been looked at with X-ray observatories since then.

There is very little information available on HIP 19265 and HIP 88580, other than what has been given in Table 1,
except that they may have infrared excesses (McDonald, Zijlstra, \& Boyer 2012).
 
Capella, apart from being a very bright visible star system, believed to be a spectroscopic binary consisting of
a K0 III star plus a rapidly rotating (period$\sim$8 days) G1 III star in a 104 day orbit 
(Hummel et al. 1994; Strassmeier \& Fekel 1990).
Both the stars are bright X-ray emitting coronal stars 
(Ayres, Schiffer, \& Linsky 1983; Linsky et al. 1998; Brickhouse et al. 2000; Gu et al. 2006; Raassen \& Kaastra 2007).

HIP 23309 is a high proper motion star and a member of the $\beta$\,Pictoris moving group (Mamjek and Bell 2014).
Anomalous high proper motion in nearby stars in $Hipparcos$ and $Gaia$ catalogs are likely to be a signature of possible substellar companions, and therefore important targets for further studies (Kervella et al. 2019).
It has rotational velocity, v sini = 5.8$\pm$1.5 km s$^{-1}$, and is very young with age estimated to be 
10$\pm$3 Myr (Weise et al. 2010). 
It was detected in soft X-rays in the ROSAT All Sky Survey II (RASS II) with a flux of 
2.30$\times$10$^{-12}$ ergs cm$^{-2}$ s$^{-1}$ in the 0.5-2.0 keV energy band (Schwope et al. 2000). 
Photometric variability in optical has been reported from this star by Kiraga (2012).

\begin{table*}
	\centering
	\caption{Log of Observations}
	\label{tab:table1}
	\resizebox{2\columnwidth}{!}{
	\begin{tabular}{lcccll} % 6 columns, alignment for each
		\hline
	Star Name & Observation ID & Start Time (UT) & Stop Time (UT) & Effective & Count Rate\\
                     &  & Y:M:D:H:M:S & Y:M:D:H:M:S & Exposure (s) & 0.3$-$3.0 keV \\
		\hline	
%    HR 2047   & 9000000074 & 2015:11:04:11:43:05 & 2015:11:04:14:18:24 & 8794 & TBD$\pm$TBD\\
    HIP 19265 & 9000000076 & 2015:11:04:15:17:59 & 2015:11:05:03:14:24 & 9296 & $<$0.01\\
    HIP 88580 & 9000000266 & 2016:01:12:17:52:39 & 2016:01:13:03:42:12 & 1947 & $<$0.02\\
    Capella   & 9000000298 & 2016:01:27:16:15:54 & 2016:01:28:23:41:12 & 30720 & 0.66$\pm$0.005\\
    HIP 23309 & 9000001720 & 2017:11:24:02:08:24 & 2017:11:24:20:27:07 & 17970 & 0.069$\pm$0.003\\ 
		\hline
	\end{tabular}}
\end{table*}

\vspace{-1em}
\section{Observations}
All observations were carried out in the photon counting (PC) mode of the SXT.  Each source was observed continuously in 
an orbit of the satellite keeping the Sun avoidance angle $\geq$ 45$^\circ$ and RAM angle 
(the angle between the payload axis to the velocity vector direction of the spacecraft) $>$ 12$^\circ$ to ensure 
the safety of the mirrors and the detector.  A log of the observations is given in Table 2.   
Level 1 Data from individual orbits received at the SXT POC (Payload Operation Centre) from 
the ISSDC (Indian Space Science Data Center) were further processed with the sxtpipeline available in the SXT 
software (AS1SXTLevel2,  version 1.4b).  The source events were calibrated, extracting Level-2 cleaned event files 
for the individual orbits were extracted. The cleaned event files of the individual orbits were merged into a 
single cleaned event file to avoid the time-overlaps in the events from consecutive orbits 
using $Julia$ based merger tool. 
The XSELECT (V2.4d) package built-in HEAsoft was used to extract the images, spectra and to examine 
light curves from the processed Level-2 cleaned event files. The useful exposure times for each source thus obtained are
listed in Table 2.  

\vspace{-1em}
\section{Data Analysis and Results} 
\subsection{Extracting X-ray events}
X-ray images were extracted from the observations of the stars shown in Table 1 and 2.
These images, extracted in the energy band of 0.3-3.0 keV, are shown in Figures 2, 3, 4 and 5 for HIP 19265, HIP 88580, 
Capella, and HIP 23309. There were no events detected from the position of HIP 19265, showing lack of any signal from 
visible light or soft X-rays. Several events were registered in the SXT data from the position of HIP 88580 
showing a strong detection as can be seen in Fig 3 (left panel). 
This star, like most A-stars, is not expected to show any X-ray emission.
An examination of the pulse-height information shown in Figure 6 (left panel) showed that almost all 
these events were confined to pulse-heights corresponding to energies below 0.7 keV and 
resembled split events (events with charge split onto neighbouring pixels, 
known as events with grades $>$1) (see Burrows et al. 2005).  These grades are used to 
distinguish between X-ray photons and charged particles (and Compton-scattered high energy photons) in CCD based cameras.
These grades can range from 0 type (single pixel events where the X-ray photon is absorbed in a single pixel of the CCD) 
to 36 types depending on the pattern of the charge splitting registered in the CCD.
Grades from 0-12 only are identified as due to X-rays, while grade zero events are generally pure X-ray events.
The default setting in the processing to level 2 data is to use grades 0-12 to maximise the number of events registered and thus improve the signal-to-noise ratio as most X-ray sources are weak. This default selection of events led to the signal
seen in the Fig.3 (left panel) and Fig 6 (red data points in the left panel). 
Selection of events with different grades can be made while using XSELECT. 
We found that by selecting events with grade 0 (single pixel events) the source practically disappeared, 
as can be seen in the image shown in Fig 3 (right panel) and Fig. 6 (black data point in the left panel). 
The few single pixel events correspond to energies below 0.3 keV, the lowest recommended threshold for the SXT.
It, therefore, appears that leaked optical photons resemble higher grades ($\geq$ 1) and are most likely 
due to arrival of several visible light photons within the readout time of 2.3775s of the CCD.  

The same technique applied to an extremely bright star like Capella, however, is not sufficient to get X-ray events.
The pile up is extremely large in the low pulse-height channels that it overwhelms the CCD electronics leading to overflow 
and registering zero counts in the centre portion leading to a dark patch shown in Figure 4, irrespective of the grades chosen.
In this case, we excluded central dark patch extending to a radius of 8 arcmin. The X-ray events were then extracted from an annular region 
with radii of 8 arcmin and 16 arcmin.  A comparison of the spectrum from such events extracted for all grades 0-12 and grade 0 is shown in 
the middle panel of Figure 6, which shows that there is a pile-up of events with all grades 0-12 even after the exclusion of 
central portion, which almost disappears for single pixel events.  
In such cases, a combination of avoiding the central region and extracting only single pixel events can work quite 
well to extract X-ray events.  This is further corroborated by the modeling of X-ray spectra thus obtained, as described below.

The fourth star in our sample, HIP 23309, is only moderately bright, and a comparison of images extracted for the grades 0-12 
and grade 0 is shown in Figure 5, while the corresponding spectra for such events recorded are shown in the right panel of Fig 6.
There seems to be a pileup for all grades but is confined below our low threshold of 0.3 keV, while the spectra above that energy
are almost identical.  Modelling of the X-ray spectra from single pixel events recorded from Capella and HIP 23309 is described below.  

\begin{figure}[!thbp] 
%Single column figure
\begin{center}
\includegraphics[width=0.7\columnwidth]{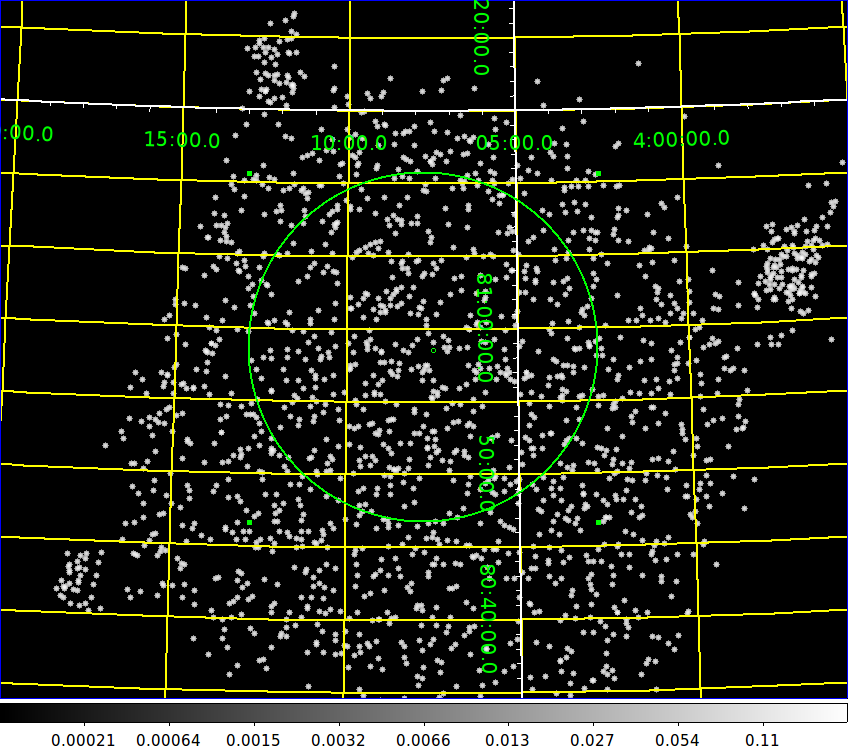}
\caption{SXT image of HIP 19265 in 0.3-3.0 keV energy band for event grades 0 to 12. 
The source region is shown as a circle centered on the position of the star and used for extraction of photons used in the analysis.}
\label{figTwo}
\end{center}
\end{figure}

\begin{figure*}
\begin{center}
\includegraphics[width=0.36\textwidth,clip]{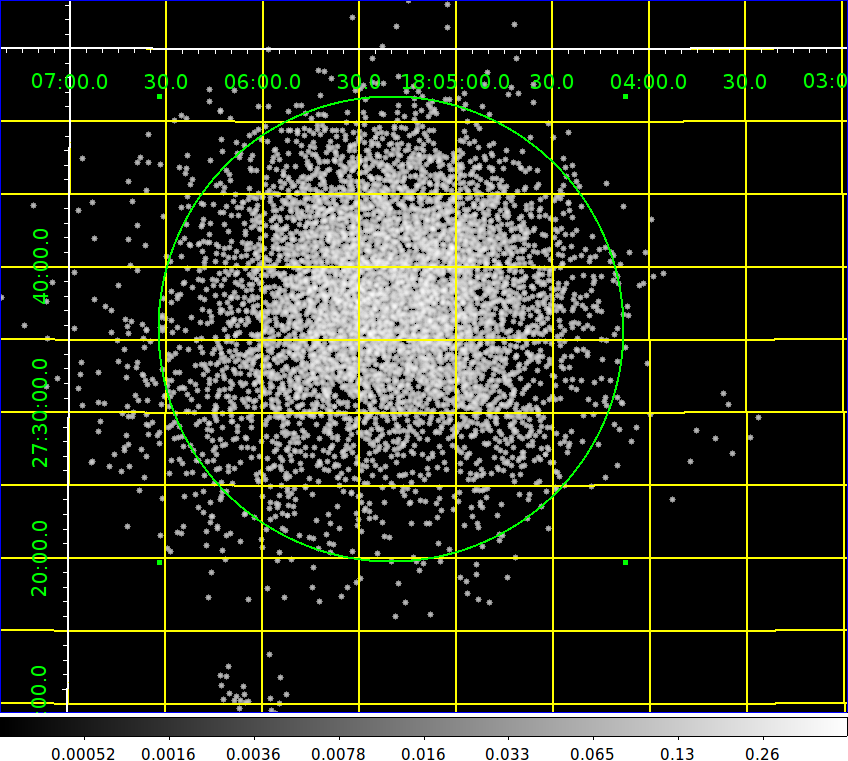}
\includegraphics[width=0.36\textwidth,clip]{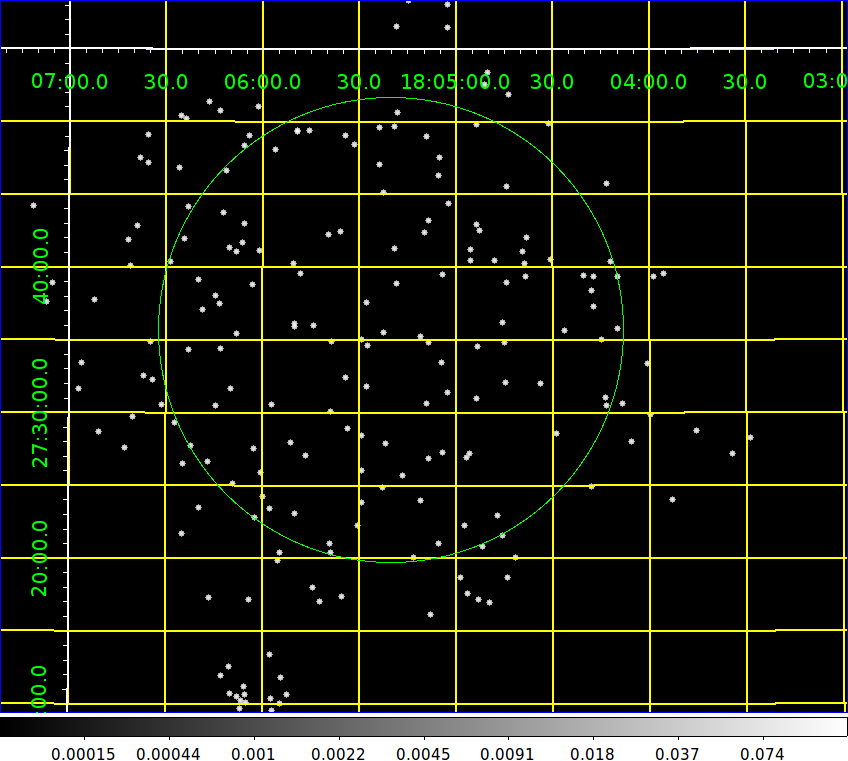}
\caption{\label{img} SXT images of HIP 88580 in 0.3-3.0 keV energy band grade 0 to 12 on the left, and grade 0 only on the right. 
The green circle shows the extraction region for getting the spectra of HIP 88580.}
\end{center}
\end{figure*}

\begin{figure*}
\begin{center}
\includegraphics[width=0.365\textwidth,clip]{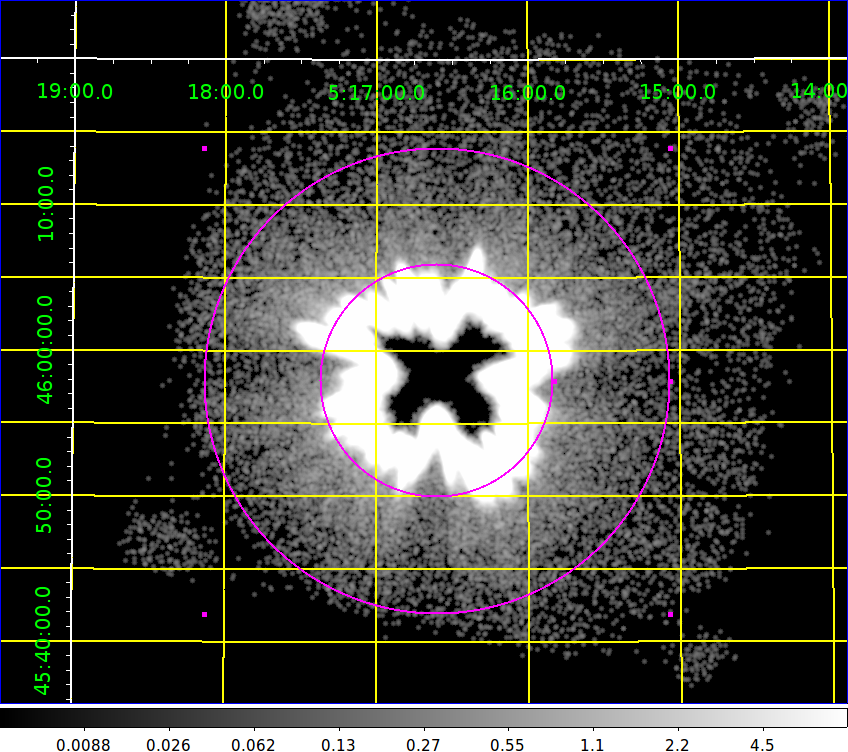}
\includegraphics[width=0.36\textwidth,clip]{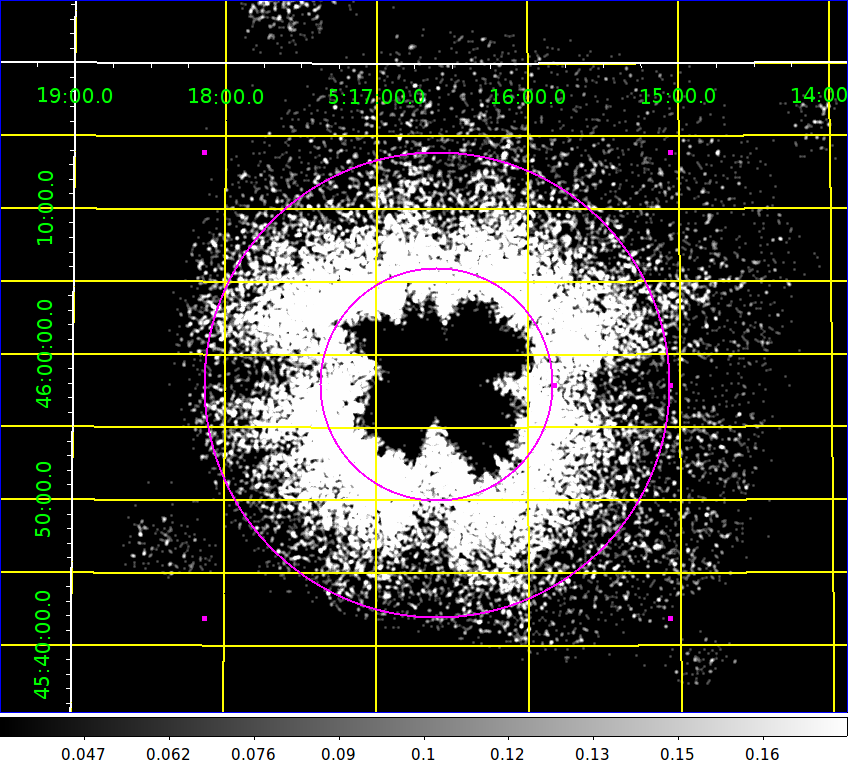}
\caption{\label{img} SXT images of Capella in 0.3-3.0 keV energy band grade 0 to 12 on the left, and grade 0 only on the right. 
The magenta circles define the annular extraction region used for getting the spectra of capella.}
\end{center}
\end{figure*}

\begin{figure*}
\begin{center}
\includegraphics[width=0.36\textwidth,clip]{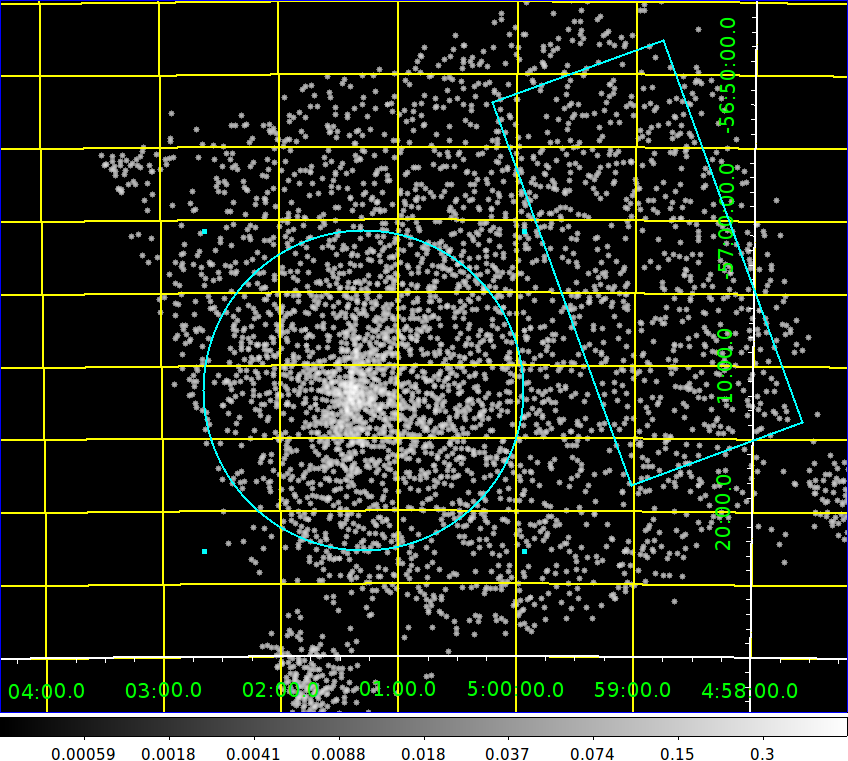}
\includegraphics[width=0.36\textwidth,clip]{hip23309_g0to12_ph30-300.png}
\caption{\label{img} SXT images of HIP 23309 in 0.3-3.0 keV energy band grade 0 to 12 on the left, and grade 0 only on the right. 
The cyan circle defines extraction region used for getting the spectra of HIP 23309, while the cyan rectangular 
box defines the extraction region for the background.}
\end{center}
\end{figure*}

\begin{figure*}
\begin{center}
\includegraphics[width=0.238\textwidth,angle=270,clip]{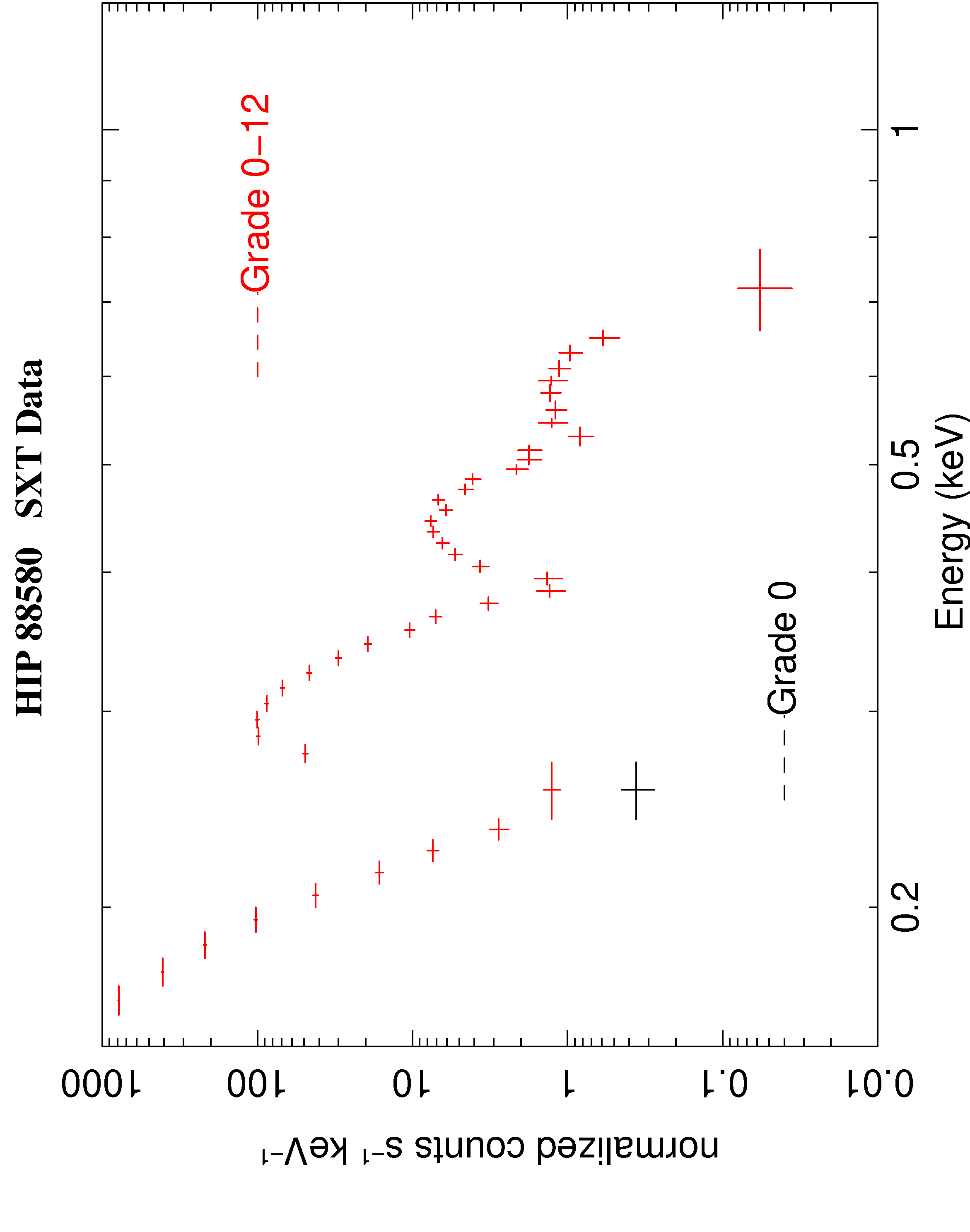}
\includegraphics[width=0.238\textwidth,angle=270,clip]{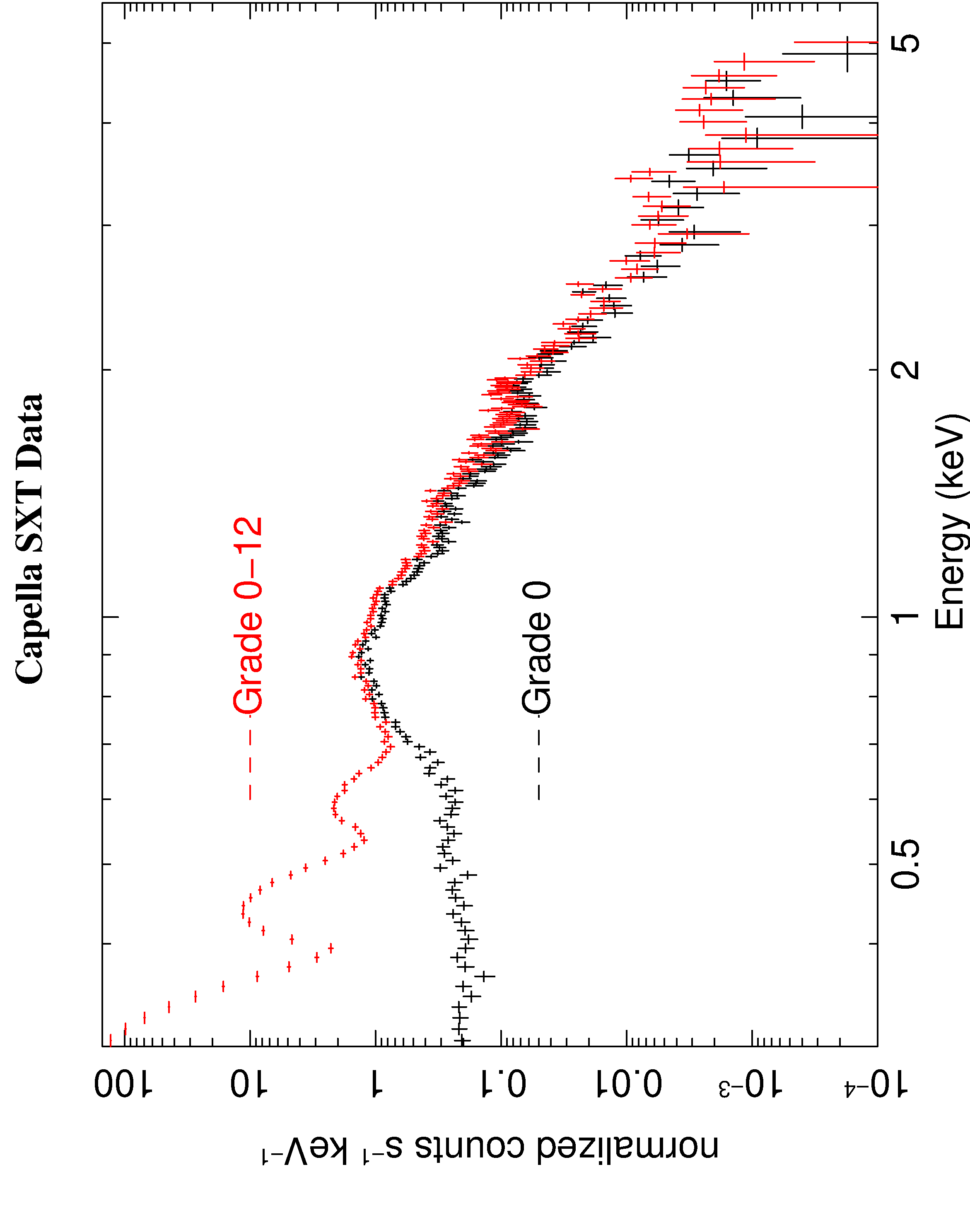}
\includegraphics[width=0.238\textwidth,angle=270,clip]{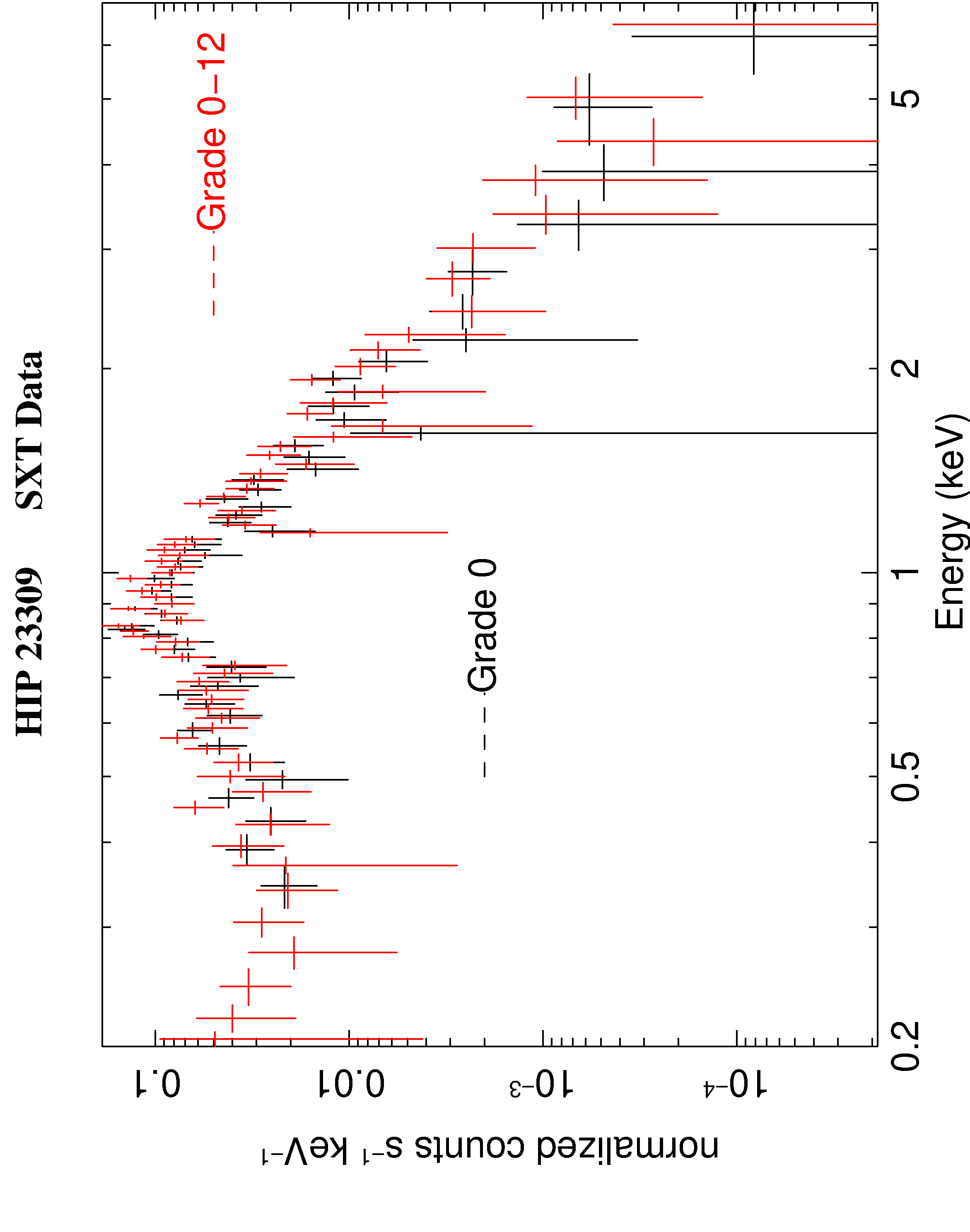}
\caption{\label{img}Comparison of spectra extracted using events with grade 0 and events with all grades from 0 to 12: HIP 88580(left), Capella (middle), HIP 23309 (right)}
\end{center}
\end{figure*}

\subsection{Modeling of X-ray spectra}
X-ray spectra were extracted for the entire observation as described above (grade 0 only) for the four stars in our sample, 
after checking that there are no variations in the count rates. The useful exposure times and the average count rates for the sources 
are given in Table 2.  The X-ray counts from the sources in the spectra were grouped using the $grppha$ tool to ensure a minimum 
of 25 counts per bin, prior to further analysis here and below.  
The response matrix, sxt\_pc\_mat\_g0.rmf, calculated for only single pixel events was used, and is 
available at the SXT POC website $https://www.tifr.res.in/~astrosat\_sxt/index.html$.
For Capella, we used a specially made ancillary response file (ARF) ”sxt\_pc\_excl00\_v04\_ann8to16arcm\_20190608.arf" 
made by using the tool $sxtmkarf$ appropriate for the source location and the annular extraction size on the CCD plane. 
For HIP 23309, we used the standard ARF file ${sxt\_arf\_excl00\_v04\_20190608.arf}$ available at the SXT POC website.
The background for the Capella was estimated from a deep exposure of 123900s on a source free region with observation ID of 9000000298, 
and using the grade 0 events only.  The background file name is "bg\_id190\_12am\_g0.pha", and this will 
be made available to public 
from the SXT POC website.  The background for the HIP 23309 was extracted from a rectangular box region of the same observation 
as the source, shown in Fig 5. Single pixel events were used here, as well.  
The source extraction region for the HIP 23309 was circular with a radius of 11 arcmin.  
The single pixel spectra for the two stars are shown in Figure 7.

X-ray spectra of Capella and HIP 23309, were fitted with optically-thin plasma emission models $apec$ using $xspec$ program 
(version 12.9.1; Arnaud (1996) distributed with the heasoft package (version 6.20). 
The atomic data base  used was AtomDB version 3.0.7 ($http://www.atomdb.org$).  
An absorber model $Tbabs$ was used as a multiplicative model with the model parameter N$_H$ , i.e., 
the equivalent Galactic neutral hydrogen column density, which was fixed at a low value of 2$\times$10$^{18}$cm$^{-2}$. 
The elemental abundance table  $"aspl"$  given by Asplund et al. (2009) was used in our analysis. 
We used $\chi^2$ minimisation technique to find the best fit 
parameters of the plasma emission models.  We tried single temperature $apec$ as well as two temperature $apec$ models, with solar 
as well as non-solar elemental abundances. The normalisation and temperature (kT) for the plasma component(s) were kept free.
The abundances of all the elements were tied together and could be varied together with respect to the solar values as one parameter.  
The results of our modelling are presented in Table 3 for Capella and Table 4 for HIP 23309.

Single temperature plasma models with solar abundances did not fit the spectrum of Capella as the reduced $\chi^2$, henceforth $\chi^2_{\nu}$, was
unacceptably high.  Similarly, two temperature models with solar abundances also gave a poor fit with  $\chi^2_{\nu}$ of 
1.827 for 188 degrees of freedom.  The fits were considerable improved when the elemental abudances were varied, either together
for all the elements or individually based on the $vapec$ models (see Table 3). The best fit was obtained with two temperature $vapec$ models
with temperature of 0.73$^{+0.014}_{-0.015}$ keV and 1.95$^{+0.70}_{-0.40}$ keV and with abundances of 
%O=0.83$^{+0.23}_{-0.21}$, Ne=0.98$^{+0.26}_{-0.25}$, Mg=0.54$^{+0.10}_{-0.09}$, Si=0.44$^{+0.09}_{-0.08}$, S=0.80$^{+0.34}_{-0.31}$, Fe=0.39$^{+0.040}_{-0.034}$; $^f$ 
O=0.94$^{+0.25}_{-0.30}$, Ne=1.25$\pm$0.35, Mg=0.55$\pm$0.11, Si=0.43$\pm$0.11, S=0.55$\pm$0.35, Fe=0.50$^{+0.07}_{-0.06}$, 
relative to the solar values.
The emission measures (EM) of the two components obtained from this best fit are 1.4$\times$10$^{53}$ 
and 2.52$\times$10$^{52}$  for the low and high temperature components, respectively. 
The best fit models are shown as histograms in Fig. 7 (left panel). 
The contributions of the two temperature components are also shown individually.  
The low temperature component dominates the emission in the best fit model. 

The X-ray spectrum of HIP 23309 could not be fitted with single temperature solar abundance plasma models.
Varying the abundance of all the elements to a very low sub-solar values gave an acceptable fit 
for a single temperature plasma.
Two temperature plasma models with solar abundances were also able to fit the data as shown by acceptable values
of the $\chi^2_{\nu}$ shown in Table 4, which improved further with sub-solar abundances.
The best fit with the lowest $\chi^2_{\nu}$ was, however, obtained by varying the abundances of the individual elements and using 
a single temperature models.  We estimate the elemental abundances in the optically-thin coronal plasma as: 
O=2.32$^{+2.7}_{-1.6}$, Ne$<$1.0, Mg$<$0.5, Si=0.4$^{+0.5}_{-0.3}$, S$<$3.0, Fe=0.2$\pm0.1$ times solar.
This best fit model is shown as a histogram in Fig. 7 (right panel).

%% use tabular font for a smaller size font
\vspace{1em}
\begin{table*}[htb]
\tabularfont
\caption{Spectral Parameters for Capella obtained from SXT data (0.3 - 5.0 keV).}\label{Spectral Paramaters} %%10/12
\begin{tabular}{lllllllll}
\topline
Spectral Model & Parameters               &            &          &              &        &      &   \\\midline
               & kT$_1$$_{apec}$         & Z$^b$      &  A$_1$$^c$  & kT$_2$$_{apec}$ & A$_2$$^c$ & $\chi^2_\nu$/dof & Flux$^d$ \\
               & keV                      &            &          &              &        &      &         \\\midline
 tbabs*apec    & 0.78                     & 1.0        &  4.61   &       -      &    -   &  3.79/190   & 1.10 \\
 tbabs*apec    & 0.78$^{+0.008}_{-0.008}$   & 0.32$^{+0.02}_{-0.02}$ & 11.3 &   -   &   -    & 1.596/189     & 1.20  \\
 tbabs*(apec+apec) & 0.73$^{+0.014}_{-0.014}$ & 1.0                  & 3.79 & 1.625$^{+0.12}_{-0.14}$ & 1.89  & 1.827/188  & 1.20 \\
 tbabs*(apec+apec) & 0.69$^{+0.04}_{-0.03}$ & 0.44$^{+0.06}_{-0.04}$ & 6.5  & 1.14$^{+0.25}_{-0.10}$ & 3.19 &  1.229/187 & 1.21 \\
 tbabs*vapec       & 0.75$^{+0.013}_{-0.013}$  & $^e$   &   8.59   &   -  &  -    &  1.358/184  &  1.21 \\
 tbabs*(vapec+vapec) & 0.73$^{+0.014}_{-0.015}$  & $^f$   & 6.66   & 1.95$^{+0.70}_{-0.40}$ & 1.37  &  1.115/182  &  1.21 \\
\hline
\hline
\end{tabular}
\tablenotes{$^a$ N$_H$ in tbabs is kept fixed at 2$\times$10$^{18}$ cm$^{-2}$ for all spectral models; $^b$ Abundance, Z,  is relative to solar values for all the elements; $^c$A$_1$ and A$_2$ are normalisations in units of 10$^{-2}$ photons cm$^2$s$^{-1}$;   Fluxes are in units of 10$^{-10}$ ergs cm$^2$ s$^{-1}$ and are quoted for total energy of 0.3 - 5.0 keV; $^e$ O=0.83$^{+0.23}_{-0.21}$, Ne=0.98$^{+0.26}_{-0.25}$, Mg=0.54$^{+0.10}_{-0.09}$, Si=0.44$^{+0.09}_{-0.08}$, S=0.80$^{+0.34}_{-0.31}$, Fe=0.39$^{+0.040}_{-0.034}$; $^f$ O=0.94$^{+0.25}_{-0.30}$, Ne=1.25$\pm$0.35, Mg=0.55$\pm$0.11, Si=0.43$\pm$0.11, S=0.55$\pm$0.35, Fe=0.50$^{+0.07}_{-0.06}$; All errors quoted are with 90\% confidence.}
\end{table*}

\begin{figure*}
\begin{center}
\includegraphics[width=0.36\textwidth,angle=270,clip]{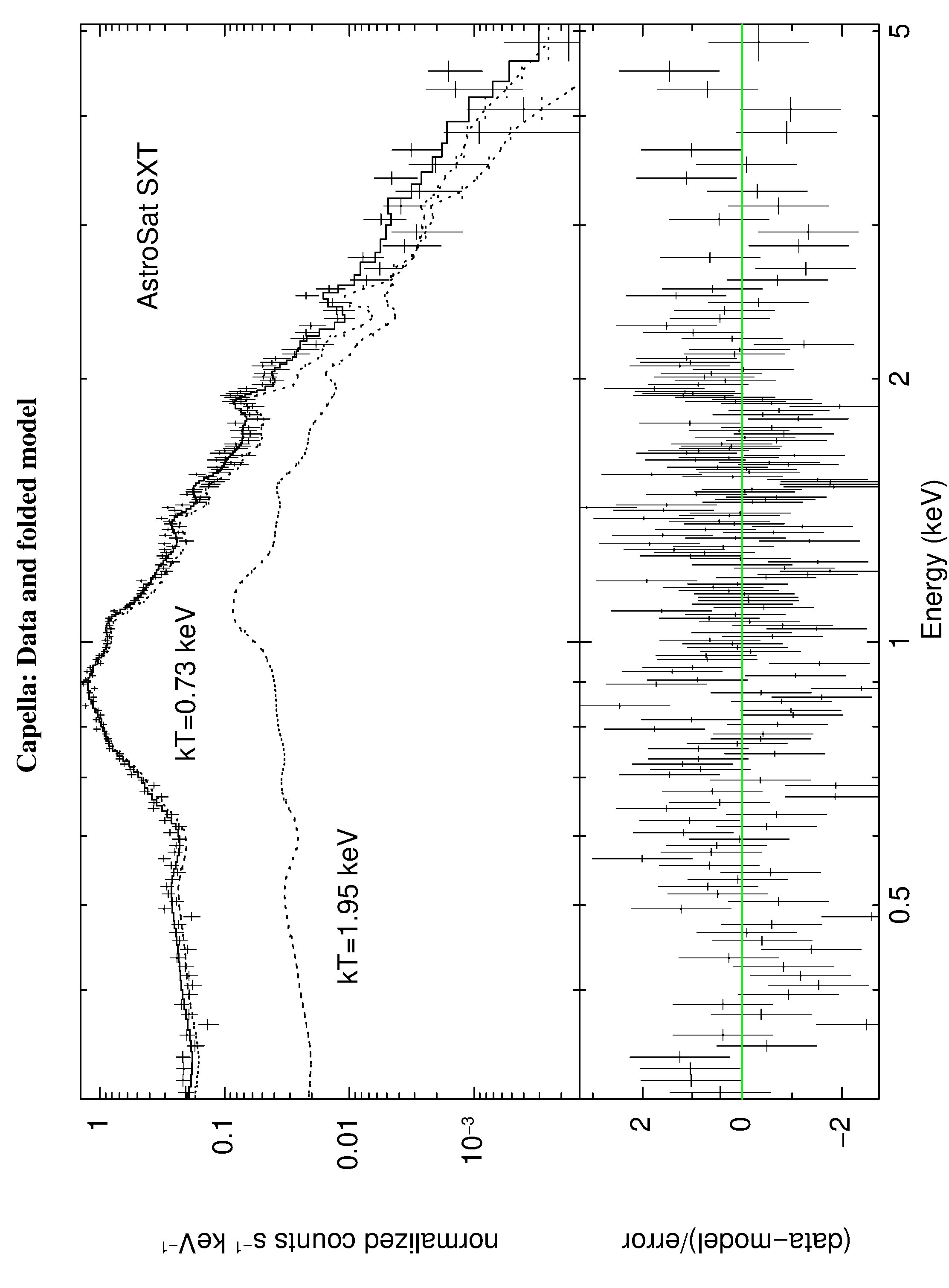}
\includegraphics[width=0.36\textwidth,angle=270,clip]{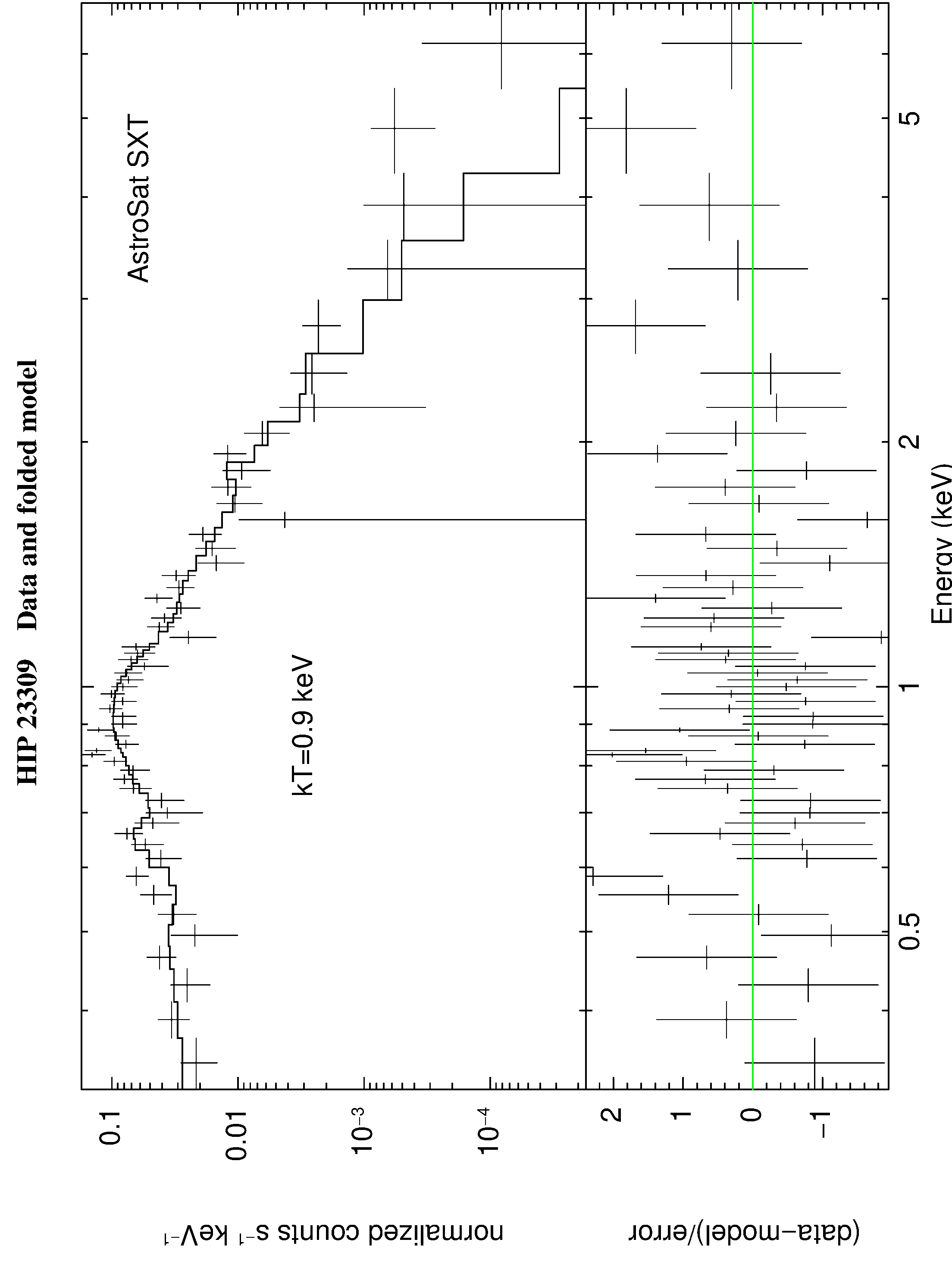}
\caption{\label{img} X-ray spectra of Capella (left) and HIP 23309 (right) with best fit optically-thin plasma models (vapec) with variable abundances.}
\end{center}
\end{figure*}

\vspace{1em}
\begin{table*}[htb]
\tabularfont
\caption{Spectral Parameters for HIP 23309 obtained from SXT data (0.3 - 7.0 keV).}\label{Spectral Paramaters} %%10/12
\begin{tabular}{lllllllll}
\topline
Spectral Model & Parameters               &            &          &              &        &      &   \\\midline
               & kT$_1$$_{apec}$          & Z$^b$      &  A$_1$$^c$  & kT$_2$$_{apec}$ & A$_2$$^c$ & $\chi^2_\nu$/dof & Flux$^d$ \\
               & keV                      &            &          &              &        &      &         \\\midline
 tbabs*apec    & 0.84                     & 1.0        &   1.18   &       -      &    -   &  2.247/56    & 2.8 \\
 tbabs*apec    & 0.90$^{+0.06}_{-0.08}$   & 0.13$^{+0.06}_{-0.04}$ & 5.30 &   -   &   -    & 1.040/55     & 3.8  \\
 tbabs*(apec+apec) & 0.79$^{+0.05}_{-0.06}$ & 1.0                & 0.85  &  3.1$^{+3.0}_{-1.0}$ & 1.39  & 1.213/54  & 4.0   \\
 tbabs*(apec+apec) & 0.83$^{+0.09}_{-0.06}$ & 0.15$^{+0.11}_{-0.05}$ & 4.5  & $\geq$1.6 & 0.86 &  0.98/53 & 4.3 \\
 tbabs*vapec       & 0.92$^{+0.06}_{-0.1}$  & $^e$   &    3.3     &   -  &  -    &  0.948/50  &  4.0 \\
\hline
\hline
\end{tabular}
\tablenotes{$^a$ N$_H$ in tbabs is kept fixed at 2$\times$10$^{18}$ cm$^{-2}$ for all spectral models; $^b$ Abundance, Z,  is relative to solar values for all the elements; $^c$A$_1$ and A$_2$ are normalisations in units of 10$^{-3}$ photons cm$^2$s$^{-1}$;   Fluxes are in units of 10$^{-12}$ ergs cm$^2$ s$^{-1}$ and are quoted for total energy of 0.3 - 7.1 keV; $^e$ O=2.32$^{+2.7}_{-1.6}$, Ne$<$1.0, Mg$<$0.5, Si=0.4$^{+0.5}_{-0.3}$, S$<$3.0, Fe=0.2$\pm0.1$; All errors quoted are with 90\% confidence.}
\end{table*}

\vspace{-1em}
\section{Discussion}
Capella has been studied quite extensively with a similarly low resolution CCD in the $ASCA$ observatory (Brickhouse et al. 2000)
and also with very high spectral resolution instruments like Low Energy Transmission Grating (LETG) and 
High Energy Transmission Gratings (HETG) aboard $Chandra$ X-ray Observatory
(Gu et al. 2006; Raassen \& Kaastra 2007).  It is known to have a very complex X-ray spectrum that has been used to refine atomic data
used in the plasma codes and also shows long term variations of $\sim$30-50\% 
(Brickhouse et al 2000; Gu et al. 2006; Raassen \& Kaastra 2007; Gu et al. 2020). In almost all these studies, one sees a continuous distribution
of emission measures with a range of temperatures from very low (kT=0.3 keV) to very high (kT=4 keV) 
(Gu et al. 2006; Raassen \& Kaastra 2007), with two peaks: one at
kT$\sim$0.55-0.70 keV and another broader peak at kT$\sim$ 1.7-2.2 keV. The best fit two temperature $vapec$ model obtained here 
has temperatures very close to these values.  The EM value of 6.7$\times$4$\pi$D$^2\times$10$^{12}$cm$^3$, where D is the distance of Capella, for the low temperature component is comparable to the peak value obtained by Gu et al (2006).  
The EM value of 1.4$\times$4$\pi$D$^2\times$10$^{12}$cm$^3$, for the high temperature component is lower than the peak value 
obtained by Gu et al (2006).  It should, however, be noted that we have used two discrete temperature components and not 
the differential emission measure (DEM) analysis adopted by Gu et al.(2006) and these values are very much dependent on the 
atomic database used and the abundances thus derived.  Our low resolution spectrum is not sufficient to carry out DEM analysis here.

We provide the first detailed spectroscopy of HIP 23309, an M0Ve star.  The X-ray flux measured by us in the energy band of 0.5-2.0 keV
of the ROSAT is 3.0$\times$10$^{-12}$ergs cm$^{-2}$ s$^{-1}$, which is about 30\% higher than the value given in the RASS II.
We measure the X-ray emission measure of the star as 2.9$\times$10$^{52}$cm$^3$ and its 
X-ray luminosity as 3.5$\times$10$^{30}$ergs s$^{-1}$ in the energy band of 0.3-7.1 keV, for the adopted distance of 26.9 pc.
These values firmly place this star as a group of extremely active M dwarfs (Singh et al. 1999), which could be the result of its 
young age and possibly a very rapid rotation.

\vspace{-1em}
\section{Conclusions}
We have shown how using single pixel events from the data recorded in the SXT observations of moderately bright stars of V$\sim$8 mag 
can be used to extract X-ray spectral information above the low threshold of 0.3 keV, despite the leakage of visibly light
photons through the thin filter of the SXT.  For stars that are extrmely bright, like Capella, one needs to disregard the photons 
from the central core of the point spread function of the SXT as well while using the single pixel event data. X-ray spectra of Capella 
and HIP 23309 have been thus extracted reliably as compared with the past measurements.  
In the process, we have provided the first detailed X-ray spectrum of a nearby young active M dwarf.  

\vspace{-1em}
%%Use section* for acknowledgements
\section*{Acknowledgements}
We thank the Indian Space Research Organisation for scheduling the observations and the Indian Space Science Data Centre (ISSDC) for making the data available. This work has been performed utilizing the calibration data-bases and auxillary analysis tools developed, maintained and distributed by AstroSat-SXT team with members from various institutions in India and abroad and the  SXT Payload Operation Center (POC) at the TIFR, Mumbai for the pipeline reduction.The work has also made use of software, and/or web tools obtained from NASA's High Energy Astrophysics Science Archive Research Center (HEASARC), a service of the Goddard Space Flight Center and the Smithsonian Astrophysical Observatory.
\vspace{-1em}

%%use \balance somewhere in the left column of the last page to balance the two columns in the end page

%%References section

\end{document}